\documentclass[12pt]{article}
\usepackage{amsmath}
\usepackage{graphicx,psfrag,epsf}
\usepackage{enumerate}
\usepackage{natbib}
\usepackage{microtype}
\usepackage{url} % not crucial - just used below for the URL 

\newcommand{\blind}{0}

\usepackage{setspace,mathtools,amssymb,siunitx,booktabs,kpfonts,xcolor}
\usepackage[font=itshape]{quoting}
\usepackage{amssymb,multirow,amsmath,graphicx,graphics,array,lscape,rotating,color,amsthm,ifsym, colortbl, courier,listings,enumitem,hyperref,wrapfig}
\usepackage{algorithm}
\usepackage{comment}
\usepackage{algpseudocode}
\usepackage{bm}
\definecolor{dkgreen}{rgb}{0,0.6,0}
\definecolor{lightgreen}{rgb}{0.7,0.95,0.7}
\definecolor{gray}{rgb}{0.9,0.9,0.9}
\definecolor{mauve}{rgb}{0.58,0,0.82}
\definecolor{lightred}{rgb}{0.95,0.8,0.8}
\definecolor{lblue}{rgb}{0.2,0.45,0.6}
\usepackage{bbm}
\usepackage{makecell}

\begin{document}

\def\spacingset#1{\renewcommand{\baselinestretch}%
{#1}\small\normalsize} \spacingset{1}

%%%%%%%%%%%%%%%%%%%%%%%%%%%%%%%%%%%%%%%%%%%%%%%%%%%%%%%%%%%%%%%%%%%%%%%%%%%%%%

\if0\blind
{
  \title{\bf Model combinations through revised base-rates}
  \author{Fotios Petropoulos\\
    School of Management,\\
    University of Bath UK\\
    and \\
    Evangelos Spiliotis \\
    Forecasting and Strategy Unit,\\
    School of Electrical and Computer Engineering,\\
    National Technical University of Athens, Greece\\
    and\\
    Anastasios Panagiotelis\\
    Discipline of Business Analytics,\\
    University of Sydney, Australia}
  \maketitle
} \fi

\if1\blind
{
  \bigskip
  \bigskip
  \bigskip
  \begin{center}
    {\LARGE\bf Model combinations through revised base-rates}
\end{center}
  \medskip
} \fi

\bigskip
\begin{abstract}
Standard selection criteria for forecasting models focus on information that is calculated for each series independently, disregarding the general tendencies and performances of the candidate models. In this paper, we propose a new way to statistical model selection and model combination that incorporates the base-rates of the candidate forecasting models, which are then revised so that the per-series information is taken into account. We examine two schemes that are based on the precision and sensitivity information from the contingency table of the base rates. We apply our approach on pools of exponential smoothing models and a large number of real time series and we show that our schemes work better than standard statistical benchmarks. We discuss the connection of our approach to other cross-learning approaches and offer insights regarding implications for theory and practice.
\end{abstract}

\noindent%
\vfill

\newpage
\spacingset{1.7} % DON'T change the spacing!
\section{Introduction}

\begin{sloppypar}

Model selection and combination have long been fundamental ideas in forecasting for business and economics \citep[see][and references therein for model selection and combination respectively]{InoKil2006,Tim2006}.  In both research and practice, selection and/or the combination weight of a forecasting model are \textit{case-specific}. By this, we mean that they are based on criteria such Akaike's information criterion \citep{Kol2011} or predictive log score \citep{GewAmi2011, PetTim2017} computed only on the \textit{series of interest itself}. This neglects important base-rate, environmental information, in particular the propensity across a wide range of time series for a selection criterion to favour a model with poor out-of-sample forecasting performance. 

\end{sloppypar}
%This resembles an example considered by \cite{Hoch1996-ta} who in reviewing psychological approaches in decision making write
%\begin{quoting}
%	Tversky and Kahneman argue that people often make categorisation judgments (e.g., how likely is it that object A belongs to category B) according to how well the current situation matches the stereotype of a category stored in memory. When making judgments of this type people should pay attention to both general tendencies in the environment (called base-rate information) and features of the current situation (called case-specific information). Numerous papers demonstrate, however, that people over-emphasize case-specific information (i.e., the pattern) at the expense of base-rate information, which they often ignore completely.
%\end{quoting}

In this paper we propose easy to implement and general algorithms for model selection and combination in forecasting that exploit revised base-rate information by using a large collection of reference series. Examples of reference series could include the large collections of macroeconomic time series \citep{StoWat2012}, or the time series from the M forecasting competitions\cite{Makridakis2020-mm}. Rather than use these reference series as predictors, they are instead used to revise the probabilities that a model is the 'correct' model in the sense of having the best forecasting performance. Apart from the reference series, the only other requirements are the choice of a pool of candidate models, a criterion for selecting between these models, and a criterion for evaluating forecasts. As a result, there is scope to tailor our proposed algorithms to applications with specific loss functions.  Furthermore, as long as the selection and evaluation criteria are likelihood-free, the set of candidate models can even include models for which the likelihood is intractable or difficult to compute.

To provide the general idea behind our proposed approach, let there be two models under consideration (model A and model B). Let events $S_\text{A}$ and $S_\text{B}$ refer to models A or B respectively being selected according to some criterion. Similarly, let events $C_\text{A}$ and $C_\text{B}$ refer to models A or B being the ``correct'' model, in the sense of being optimal with respect to some evaluation criterion. Also, assume that we have access to a large set of reference series, such that we can empirically estimate joint probabilities of models being selected and ``correct'', and thus populate the cells of the contingency table (such as table \ref{tab:contingencytable}). The use of reference series is inspired by the meta-learning literature in forecasting \citep[see][and references therein]{LemGab2010,WanEtAl2009,TalEtAl2018,Montero-Manso2020-et}.  However, in contrast to these papers, the weights we compute have an interpretation as probabilities rather than being the outputs of a ``black-box'', machine learning algorithm which type (e.g., neural network or decision tree) and hyper-parameter values have to be carefully selected. Moreover, since the weights in our approach are solely estimated using forecasting performance related information, they are not subject to time series features and general statistics which number, type, and representativeness may be challenging to determine in practice for constructing a successful meta-learning algorithm. In addition, the information exploited by our approach focuses on models instead of series, being also summarised at a global level (forecasting performance is being tracked across the complete set of reference series) instead of being learnt at a local one (forecasting performance is being tracked at each series and the connections between the inputs and outputs of the algorithm are determined accordingly).  

From table~\ref{tab:contingencytable}, we can observe the general, environmental tendencies for models A and B in terms of (\textit{i}) being the selected model with probabilities $p(S_\text{A})$ and $p(S_\text{B})$, respectively, and (\textit{ii}) being the ``correct'' model with probabilities $p(C_\text{A})$ and $p(C_\text{B})$, respectively. In the literature, when selection is based on $p(C_\text{A})$ and $p(C_\text{B})$, which correspond to the base-rate information, it is typically called ``aggregate selection'' as a single model is used to forecast all series by considering which model provided the most accurate forecasts for a hold-out sample in most of the cases and ignoring their particular characteristics  \citep{FILDES20151692}.

\begin{table}[h]
	\centering
	\caption{The contingency table.}\vspace{0.25cm}
	\begin{tabular}{c|cc|c}
		& $C_\text{A}$ & $C_\text{B}$ & Total \\
		\hline
		$S_\text{A}$ & $p(S_\text{A} \cap C_\text{A})$  & $p(S_\text{A} \cap C_\text{B})$ &  $p(S_\text{A})$ \\
		$S_\text{B}$ & $p(S_\text{B} \cap C_\text{A})$  & $p(S_\text{B} \cap C_\text{B})$ &  $p(S_\text{B})$ \\
		\hline
		Total & $p(C_\text{A})$ & $p(C_\text{B})$ & 1
	\end{tabular}
	\label{tab:contingencytable}
\end{table}

For any new series, we can first evaluate the case-specific event of selecting either model A or model B, i.e., we either observe the event $S_\text{A}$ or $S_\text{B}$.  Suppose we observe $S_\text{A}$.  Rather than use model $\text{A}$ we propose to incorporate base-rate information by selecting model A when $p(C_\text{A}|S_\text{A})>p(C_\text{B}|S_\text{A})$ and model B otherwise. These conditional probabilities are computed using the values in the contingency table. These conditional probabilities summarise ``precision'', i.e., the proportion of cases for which the selected model is actually the ``correct'' model. An alternative approach will be to select model A when $p(S_\text{A}|C_\text{A})>p(S_\text{A}|C_\text{B})$ and model B otherwise.  These conditional probabilities offer the revised probabilities for a model being selected, assuming that some other model (or the same model) is the correct one. In contrast to the previous case, these conditional probabilities summarise  ``sensitivity''.  We note that  $p(S_\text{A}|C_\text{A})=(C_\text{A}|S_\text{A})$ and $p(S_\text{A}|C_\text{B})=(C_\text{B}|S_\text{A})$ only when $p(C_\text{A})=p(C_\text{B})$.

If instead of model selection, model averaging is desired, the conditional probabilities discussed above can be used as weights in a forecast combination.  Since $p(C|S)$ represents the probability that a model is the ``correct'' model conditional on observed information, this bears an interesting resemblance to the Bayesian paradigm for model averaging.  In the Bayesian setting, the choice of model is treated in the same way as other parameters and model averaging based on posterior model probabilities arises in a natural way to integrate model uncertainty.  The posterior probability that a given model is the ``correct'' model can be computed using Bayes theorem, although in practice Markov chain Monte Carlo algorithms are required for exploring the model space. For an extensive review of Bayesian model averaging, including key historical references, see \citet{HoeEtAl1999}. 

Although the computation of posterior model probabilities can be challenging, there are a number of useful approximations.  A prominent example, discussed by \citet{Raf1995}, is based on the Bayesian Information Criterion (BIC), which we use as one of the selection criteria in Section~\ref{sec:selection_criteria}.  Alternatives to the BIC can also be considered. For example, in the forecasting literature \citet{Kol2011} uses Akaike's Information Criterion (AIC) in a similar fashion to find forecast combination weights.  Furthermore, recent work by \citet{BisEtAl2016}, \citet{Loa2020} as well as the literature on PAC-Bayes \citep[see][for a review]{Gue2019} generalise posterior inference to allow loss functions to replace likelihoods and can also be applied to finding model weights. Our proposed approach also uses loss functions in the form of selection and evaluation criteria. It does however differ from existing approaches substantially, by using the conditional probabilities computed from reference series as ``proxies'' for the posterior model probabilities. By comparing our own proposed methods to forecast combinations that approximate posterior model probabilities without using the reference time series, we can examine the benefits of the proposed cross-learning framework. Furthermore, the use of general loss functions, rather than likelihoods allow our method to be extended to machine learning methods such as random forests which are becoming increasingly popular in business and macroeconomic forecasting \citep{MedEtAl2021}.

The remainder of the paper is structured as follows.  Section~\ref{sec:meth} introduces our proposed approach more rigorously including different approaches for computing the weights from the reference time series.  Section~\ref{sec:empiricaldesign} describes the empirical design used to evaluate our proposed approach.  Section~\ref{sec:empiricalresults} summarises the main results while Section~\ref{sec:discussion} and Section~\ref{sec:conclusions} provide additional discussion and conclude respectively. 

\section{Methodology}\label{sec:meth}

The proposed forecasting algorithm involves two steps. First, a large collection of reference time series are used to compute the probabilities in a contingency table. Second, models are fit to the actual time series of interest and combination weights are derived for each model according to one of three schemes, two of which depend on the contingency table.  We now describe each of these steps in turn.

\subsection{Populating contingency table}\label{sec:algo_ct}

Let $\mathcal{Z}:=\left\{z^{(1)},\dots,z^{(N)}\right\}$ be a collection of $N$ \textit{reference} time series relevant to, but not including, the time series of interest.  Let $\mathcal{M}:=\left\{M_{1},\dots,M_{K}\right\}$ be a set of $K$ \textit{models} that can be used for forecasting.  Let $S$ be a \textit{selection criterion}, computed using only in-sample information and with value $S^{(n)}_{k}$ for reference time series $z^{(n)}$ and model $M_{k}$. Similarly, let $C$ be an \textit{evaluation criterion}, used to determine the ``correct'' model, that is computed using only out-of-sample information and has value $C^{(n)}_{k}$ for reference time series $z^{(n)}$ and model $M_{k}$. Without loss of generality, we will assume that lower values of $S^{(n)}_{k}$ and $C^{(n)}_{k}$ indicate better performing models. Although, the selection and evaluation criteria will be of a statistical nature in this paper, in certain applications, model combinations could be based on context specific loss functions \citep[for an example from finance see][who consider model combinations based on Sharpe ratios]{CalEtAl2016}.

Let $\bm{W}$ be a $K\times K$ matrix corresponding to the contingency table, with element $w_{i,j}$ in the $i^\text{th}$ column and $j^\text{th}$ row. These entries measure the joint probability that for a randomly selected reference time series, the ``correct'' model is model $j$ when the selected model is model $i$. Algorithm~\ref{alg:pop_ct} provides details on how $w_{i,j}$ are computed.

\begin{algorithm}
	\caption{Algorithm to populate cells of contingency table}\label{alg:pop_ct}
	\begin{algorithmic}[1]
		\Procedure{ContTab}{$\mathcal{Z}$,$\mathcal{M}$,$S$,$C$}
		\State Set $w_{i,j}\leftarrow 0$ for all $i=1,\dots,K$, $j=1,\dots,K$ \Comment{Initialise}
		\For{$n=1,\ldots,N$}\Comment{Loop over reference time series}
		\State Split $z^{(n)}$ into a training sample $z^{(n)}_{train}$ and a test sample $z^{(n)}_{test}$.
		\For{$k=1,\dots,K$} \Comment{Loop over models}
		\State Fit model $M_k$ to $z^{(n)}_{train}$ and compute $S^{(n)}_{k}$.
		\State Compute $C^{(n)}_{k}$ using $z^{(n)}_{test}$
		\EndFor
		\State Set $i^*=\underset{i}{argmin}S^{(n)}_{i}$ \Comment{if larger values of $S^{(n)}_{i}$ indicate better models, use $argmax$ instead}
		\State Set $j^*=\underset{j}{argmin}C^{(n)}_{j}$ \Comment{if larger values of $C^{(n)}_{j}$ indicate better models, use $argmax$ instead}
		\State Set $w_{i^*,j^*}\leftarrow w_{i^*,j^*}+1$
		\EndFor
		\State Set $\bm{W}\leftarrow \bm{W}\big{/}N$ \Comment{Normalise cells of contingency table}
		\EndProcedure
	\end{algorithmic}
\end{algorithm}

\subsection{Forecasting algorithm}\label{sec:algo_fore}

Now let $\bm{y}$ be the time series of interest that needs to be forecast.  Forecasts from all models in $\mathcal{M}$ will be produced.  Three schemes are considered for model selection and averaging.  \textit{Criterion}-based schemes ignore the information in the contingency table entirely. \textit{Precision}-based schemes derive weights based on the probabilities that a model is the ``correct'' model conditional on it being selected ($p(C|S)$). \textit{Sensitivity}-based schemes derive weights based on probabilities of selecting a model given the ``correct'' model  ($p(S|C)$) and are equivalent to $p(C|S)$ if the assumed distribution $p(C)$ is uniform (a priori the new time series is equally likely to be best forecast by any model). The computation of weights is outlined in detail in Algorithm~\ref{alg:comp_pr}.

\begin{algorithm}
	\caption{Algorithm to compute criterion, precision and sensitivity combination weights}\label{alg:comp_pr}
	\begin{algorithmic}[1]
		\Procedure{CompW}{$\mathbf{W}$,$\mathcal{M}$,$S$,$\bm{y}$}
		\For{k=1,\dots K} \Comment{Loop over models}
		\State Fit model $M_k$ to $\bm{y}$ and compute $S^{\bm{y}}_{k}$ and a forecast $\hat{y}_{T+h,k}$.
		\EndFor
		\State Set unnormalised criterion weights to $w^{\textrm{crit}}_k\leftarrow exp(-S^{\bm{y}}_{k}/2)$
		\State Set $i^*=\underset{i}{argmin}S^{\bm{y}}_{i}$ \Comment{if larger values of $S^{\bm{y}}_{i}$ indicate better models then use $argmax$ instead}
		\State Compute unnormalised precision weights $w^{\textrm{prec}}_k\leftarrow w_{i^*,k}$
		\State Compute unnormalised sensitivity weights $w^{\textrm{sens}}_k\leftarrow w_{i^*,k}\big{/}\sum\limits_i w_{i,k}$
		\State Normalise all weights, $\bm{w}^{\textrm{g}}\leftarrow \bm{w}^{\textrm{g}}\big{/}\sum\limits_k w^{\textrm{g}}_{k}$, for $\textrm{g}\in\left\{\textrm{crit},\textrm{prec},\textrm{sens}\right\}$ and where bold $\bm{w}$ denotes vectors of length $K$.
		\EndProcedure
	\end{algorithmic}
\end{algorithm}

The final forecasts are either based on selection or averaging and one of the three weighting schemes.  In the results of Section~\ref{sec:empiricalresults}, \textit{Criterion-select}, \textit{Precision-select} and \textit{Sensitivity-select} respectively refer to using the model corresponding to the element that is maximal for $w^{\textrm{crit}}$, $w^{\textrm{prec}}$ and $w^{\textrm{sens}}$.  Alternatively,  \textit{Criterion-average}, \textit{Precision-average} and \textit{Sensitivity-average} take a weighted average of forecasts using $w^{\textrm{crit}}$, $w^{\textrm{prec}}$ and $w^{\textrm{sens}}$ respectively as weights. % As a benchmark, an equally weighted forecast combination, referred to as \textit{EQ-average}, is also considered.

\section{Empirical design}\label{sec:empiricaldesign}

\subsection{Models}\label{sec:models}

In this study, we focus on the exponential smoothing (ETS) family of models. In exponential smoothing models, up to three components are estimated (level, trend, and seasonality) and the forecasts are based on the estimates of these components. The exponential smoothing family consists of thirty models in total, which are all possible combinations of different types of error (additive or multiplicative), trend (none, additive, or multiplicative; damped or not), and seasonality (none, additive, or multiplicative). An exponential smoothing model form is usually summarised by three or four letters that represent the types of the components in the model. For instance, an exponential smoothing model with additive error, additive trend, and multiplicative seasonality is acronymised as ETS(AAM), or simply AAM. Similarly, ETS(MAdN) is a model with multiplicative error, additive damped trend, and no seasonal component.

In practice not all models are used, either because some combinations result in estimation difficulties (such as additive error term with multiplicative seasonality) or in unrealistic and explosive forecasts (such as multiplicative trends). The very popular \textit{forecast} package for the R statistical software \citep{Hyndman2020-tp}, which is used for this study, considers by default fifteen out of the thirty theoretically possible models. For non-seasonal data (such as time series with a yearly frequency), the number of available exponential smoothing models drops from fifteen to six.

\begin{sloppypar}

There are three reasons that make the use of exponential smoothing attractive and relevant for this study. First, exponential smoothing models are a mainstream option in practice \citep{Weller2012}. Second, they offer a robust performance \citep{Makridakis2000-ty} and low computational cost \citep{Makridakis2020-mm}. Third, the exponential family of models is finite, allowing for a linear search to identify an optimal model. On the other hand, the autoregressive integrated moving average (ARIMA) family of models, another very popular univariate modelling approach, consists of a theoretically infinite number of models. ARIMA implementations are usually based on non-linear stepwise-type searches across models. Even if a maximum order for ARIMA models is assumed to allow for a sequential non-stepwise search, the computational cost and the number of possible models increase significantly \citep{Petropoulos2021-ec} resulting in sparse base-rate matrices. Finally, although multivariate models could in principle be used, we prefer univariate models due to ease of implementation across a large number of reference series and note that in practice, multivariate models do not necessarily outperform univariate models in forecasting \citep[for a recent discussion see][and references therein]{MitEtAl2019}.

\end{sloppypar}

\subsection{Selection and evaluation criteria}\label{sec:selection_criteria}

We consider two selection criteria, which are described below. The evaluation criterion that we use in this study is the mean absolute error (MAE). The application of the proposed algorithm for populating the contingency table (subsection \ref{sec:algo_ct}) requires the splitting of each reference time series into a training set ($z^{(n)}_{train}$ -- on which the selection criteria values are calculated) and a test set ($z^{(n)}_{test}$ -- on which the evaluation criterion values are calculated). Let $T^{(n)}$ be the length of the reference time series $z^{(n)}$ and $h$ the required forecast horizon for $\bm{y}$. The first $T^{(n)}-h$ observations of $z^{(n)}$ serve as the training data, with the last $h$ being the test data.

The first selection criterion is an information criterion. Information criteria are based on the in-sample performance of a model, penalised for the size of the model (number of parameters that need to be estimated). Information criteria values can be calculated using the training data of each reference time series, $z^{(n)}_{train}$. In this study, we present results for the BIC. However, the insights are consistent for other information criteria such as the AIC or its corrected version for small sample sizes (AICc).

The second selection criterion is time series validation. Replacing information criteria with time series validation allows us to directly match the cost functions of the selection and the evaluation criteria in constructing the base-rate matrix. However, this requires further splitting the series such that selection via validation is enabled. We first consider the first $T^{(n)}-2h$ observations of $z^{(n)}$ and prepare forecasts for the periods $T^{(n)}-2h+1$ to $T^{(n)}-h$. The model that performs best (based on MAE) on the periods $T^{(n)}-2h+1$ to $T^{(n)}-h$ is the selection via time series validation. Next, we take the first $T^{(n)}-h$ observations of $z^{(n)}$, corresponding to $z^{(n)}_{train}$, and prepare forecasts for $z^{(n)}_{set}$ to calculate the evaluation criterion, similarly to information criteria.

\subsection{Data}\label{sec:data}

We use the yearly, quarterly, and monthly data from the M, M3, and M4 forecasting competitions \citep{Makridakis1982-co,Makridakis2000-ty,Makridakis2020-mm}. The forecasting horizon considered in this study is different per data frequency, which matches the original design of the aforementioned competitions: $h=6$, $8$, and $12$ for the yearly, quarterly, and monthly data respectively. Following the process described in \ref{sec:algo_ct}, we populate a separate contingency table per frequency and selection criterion (BIC or time series validation).

The various exponential smoothing models available have different numbers of parameters to be estimated and, as such, require a different minimum number of available observations. Because of that, we need to ensure that models are selected for their merits and not their data requirements. This would be important for the shorter of the series available, as their inclusion would introduce a bias when populating the contingency tables. As such, for each selection criterion, only the series that could be fitted over all available exponential smoothing models were considered. Table \ref{tab:series} provides the respective counts. 

Finally, the out-of-sample evaluation takes place on the series that both selection criteria can be applied, which matches the counts of series for constructing the BIC's contingency tables. That means that for the case of selection with validation, there is not an absolute match between the series used for constructing the base rate matrices and the series that are finally evaluated. This mismatch would be a normal situation for cases where only a small number of series needs to be forecasted, with the corresponding contingency table being populated using a wider, representative set of time series.

\begin{table}[h]
	\centering
	\caption{Number of series considered for constructing the base-rate matrices for each selection criterion, i.e., BIC and time series validation.}\vspace{0.25cm}
	\begin{tabular}{c|cc}
		Frequency & BIC base-rate & Validation base-rate \\
		\hline
		Yearly & 20,616 & 15,315 \\
		Quarterly & 24,820 & 24,327 \\
		Monthly & 49,998 & 49,477 \\
		\hline
		Total & 95,434 & 89,119 \\
	\end{tabular}
	\label{tab:series}
\end{table}

\subsection{Measuring performance}\label{sec:measures}

Following \cite{Makridakis2020-mm}, we consider two forecasting performance measures. The first focuses on the point forecast accuracy, while the second assesses the performance of prediction intervals. Let $y_t$ be the observation of $\bm{y}$ at time period $t$ and $f_t$, $u_t$ and $l_t$ the point forecast, the upper and the lower prediction interval for the same period, respectively. Also, let $T$ be the length of in-sample data for $\bm{y}$ and $s$ its seasonality (e.g., $s=12$ for monthly data). The Mean Absolute Scaled Error (MASE) is a widely-used measures of point forecast accuracy and is defined as
\begin{align*}
%\text{sMAPE} &= \frac{200}{h} \displaystyle\sum_{t=T+1}^{T+h} \frac{|y_{t}-f_{t}|} { |y_{t}| + |f_{t}| },\\
\text{MASE} &= \frac{1}{h} \frac{ \displaystyle\sum_{t=T+1}^{T+h} {|y_{t}-f_{t}|} } {\frac{1}{T-s} \displaystyle\sum_{t=s+1}^{T} |y_{t}-y_{t-s}|}.
\end{align*}

The Mean Scaled Interval Score (MSIS) is used as a measure of the performance of the prediction intervals. It is the scaled average difference between upper and lower prediction interval plus a penalty for the instances where the actual observation lies outside the intervals. This penalty is linked to the desired confidence level, $(1-\alpha) \times 100 \%$. In this study, we set $\alpha=0.05$ (95\% confidence level). The MSIS is defined as
$$
\text{MSIS} = \frac{1}{h} \frac{ \displaystyle\sum_{t=T+1}^{T+h} \left( u_t - l_t + \frac{2}{\alpha}(l_t - y_t)\mathbbm{1}\{y_t < l_t\} + \frac{2}{\alpha}(y_t - u_t)\mathbbm{1}\{y_t > u_t\} \right) } {\frac{1}{T-s} \displaystyle\sum_{i=s+1}^{T} |y_{t}-y_{t-s}|},
$$
\noindent in which $\mathbbm{1}\{ \cdot \}$ is an indicator function. Note that the scaling of MSIS is the same as in MASE. The values of the two measures, MASE and MSIS, can be averaged across many time series. For both measures, lower values are better. 

\section{Empirical results}\label{sec:empiricalresults}

\subsection{Contingency tables, precision and sensitivity rates, and $F$-scores}\label{sec:contabs}

The contingency tables for the yearly data are provided in tables \ref{tab:contab1} and \ref{tab:contab2} for the BIC and the Validation selection criteria respectively. When BIC is used as the selection criterion, we can see that the exponential smoothing model with multiplicative error form and additive seasonality (MAN) is selected as optimal in 44.6\% of the reference series, while it is the ``correct'' in only 24.1\% of cases. Moreover, the more complex damped-trend models (AAdN and MAdN) are selected much less often, only 3.6\% of the times, despite being the best options for more than 1/4 of the cases. Overall, we can see that BIC tends to select simpler models, which follows the logic of its construction and the penalisation applied on models with more parameters. Selection via time series validation results in a more balanced use of the available models in the pool. Similar insights are gained from the contingency tables of the seasonal series (quarterly and monthly), which are not presented for brevity.

\begin{table}[h]
	\centering
	\caption{Contingency table for the yearly data when BIC is used as the selection criterion (rows) and MAE as the evaluation criterion (columns).}\vspace{0.25cm}
	\begin{tabular}{c|cccccc|c}
		ETS Model & ANN	 & MNN	 & AAN	 & MAN	 & AAdN	 & MAdN & Total\\
		\hline
		ANN & 0.026	 & 0.019	 & 0.030	 & 0.032	 & 0.013	 & 0.013	 & 0.134 \\
		MNN & 0.038	 & 0.026	 & 0.047	 & 0.048	 & 0.022	 & 0.022	 & 0.203 \\
		AAN & 0.017	 & 0.010	 & 0.062	 & 0.041	 & 0.025	 & 0.025	 & 0.181 \\
		MAN & 0.043	 & 0.027	 & 0.140	 & 0.109	 & 0.062	 & 0.066	 & 0.446 \\
		AAdN & 0.002	 & 0.002	 & 0.003	 & 0.004	 & 0.002	 & 0.002	 & 0.015 \\
		MAdN & 0.002	 & 0.002	 & 0.006	 & 0.006	 & 0.003	 & 0.003	 & 0.021 \\
		\hline
		Total & 0.128	 & 0.085	 & 0.288	 & 0.241	 & 0.128	 & 0.131	 & 1.000 \\
	\end{tabular}
	\label{tab:contab1}
\end{table}

\begin{table}[h]
	\centering
	\caption{Contingency table for the yearly data when time series validation is used as the selection criterion (rows) and MAE as the evaluation criterion (columns).}\vspace{0.25cm}
	\begin{tabular}{c|cccccc|c}
		ETS Model & ANN	 & MNN	 & AAN	 & MAN	 & AAdN	 & MAdN & Total\\
		\hline
		ANN & 0.024 & 0.014 & 0.028 & 0.032 & 0.014 & 0.014 & 0.127 \\
		MNN & 0.015 & 0.011 & 0.024 & 0.023 & 0.009 & 0.010 & 0.092 \\
		AAN & 0.025 & 0.017 & 0.092 & 0.062 & 0.037 & 0.036 & 0.268 \\
		MAN & 0.024 & 0.015 & 0.081 & 0.065 & 0.033 & 0.035 & 0.254 \\
		AAdN & 0.013 & 0.009 & 0.040 & 0.034 & 0.017 & 0.016 & 0.129 \\
		MAdN & 0.014 & 0.008 & 0.039 & 0.037 & 0.015 & 0.016 & 0.129 \\
		\hline
		Total & 0.115 & 0.075 & 0.306 & 0.253 & 0.125 & 0.127 & 1.000 \\
	\end{tabular}
	\label{tab:contab2}
\end{table}

Next, we present the precision and sensitivity average rates as well as the $F$-score for each of the selection criteria (BIC and Validation), selection schemes (Criterion-select, Precision-select and Sensitivity-select), and frequency separately, as calculated based on the contingency tables. Precision is the ratio of the true positives (TP) by the sum of true positives and false positives (FP). Sensitivity (or recall) is the ratio of the true positives by all relevant elements (true positives and false negatives; TP $+$ FN). $F$-score is a function of the precision and sensitivity values and can be calculated as 
$$
F-\text{score} = \frac{2 \times \text{precision} \times \text{sensitivity}}{\text{precision} + \text{sensitivity}} = \frac{\text{TP}}{\text{TP} + 0.5(\text{FP}+\text{FN})}.
$$
\noindent Precision, sensitivity, and $F$-score values can be calculated for each of the available models (six for yearly data; fifteen for quarterly and monthly data). These values are then averaged across models, and the results are presented in table \ref{tab:prec_sens_f}.

\begin{table}[h]
	\centering
	\caption{Precision, sensitivity and $F$-score values averaged across models for each data frequency and selection criterion.}\vspace{0.25cm}
	\begin{tabular}{ccc|ccc}
		\multirow{2}{1.8cm}{Frequency} & Selection & Selection & \multirow{2}{1.7cm}{Precision} & \multirow{2}{1.9cm}{Sensitivity} & \multirow{2}{1.4cm}{$F$-score}\\
		& Criterion & Scheme &&&\\
		\hline
		Yearly & BIC & Criterion-select & 0.203 & 0.203 & 0.172 \\
		&& Precision-select & 0.285 & 0.283 & 0.232 \\
		&& Sensitivity-select & 0.171 & 0.163 & 0.141 \\
		\cmidrule(lr){2-6}
		& Validation & Criterion-select & 0.195 & 0.197 & 0.195 \\
		&& Precision-select & 0.300 & 0.310 & 0.304 \\
		&& Sensitivity-select & 0.202 & 0.205 & 0.203 \\
		\hline
		Quarterly & BIC & Criterion-select & 0.083 & 0.081 & 0.064 \\
		&& Precision-select & 0.139 & 0.135 & 0.102 \\
		&& Sensitivity-select & 0.084 & 0.072 & 0.058 \\
		\cmidrule(lr){2-6}
		& Validation & Criterion-select & 0.089 & 0.089 & 0.089 \\
		&& Precision-select & 0.126 & 0.126 & 0.126 \\
		&& Sensitivity-select & 0.100 & 0.101 & 0.100 \\
		\hline
		Monthly & BIC & Criterion-select & 0.086 & 0.087 & 0.068 \\
		&& Precision-select & 0.162 & 0.180 & 0.133 \\
		&& Sensitivity-select & 0.107 & 0.112 & 0.090 \\
		\cmidrule(lr){2-6}
		& Validation & Criterion-select & 0.094 & 0.094 & 0.094 \\
		&& Precision-select & 0.138 & 0.139 & 0.138 \\
		&& Sensitivity-select & 0.091 & 0.092 & 0.091 \\
		\hline
	\end{tabular}
	\label{tab:prec_sens_f}
\end{table}

We observe that while BIC has higher Criterion-select precision and sensitivity rates for the yearly data, Criterion-select with time series validation offers higher average $F$-score across all frequencies. Comparing across the selection schemes, Precision-select is outperforms by some margin Criterion-select across all three scores: precision, sensitivity, and $F$-score. For the yearly and quarterly data, Sensitivity-select is worse than Criterion-select when BIC is used as the selection criterion, but better when Validation is the selection criterion. The exact opposite is true for the monthly data. Finally, it is worth-noting that the values of precision, sensitivity, and $F$-score decrease as we move from the yearly to the seasonal (quarterly and monthly) data as a result of the increase in the number of available models (six versus fifteen). Overall, the results suggest that Precision-select is able to identify the correct model more often than either Criterion-select or Sensitivity-select (precision) but also to minimise the instances that a correct model is not selected (sensitivity).

\subsection{Out-of-sample evaluation}\label{sec:contabs}

Tables \ref{tab:outofsample_performance} presents the average values of MASE and MSIS on the out-of-sample performance for the various selection and combination schemes considered (see section \ref{sec:algo_fore}), for each selection criterion (see section \ref{sec:selection_criteria}), and for each and data frequency. As described in section \ref{sec:data}, the out-of-sample evaluation takes place over 95,434 yearly, quarterly, and monthly series for which both selection criteria (BIC and Validation) can be applied. The best performances (lower MASE or MSIS value) for each scheme (selection or combination) are highlighted in boldface. We excluded, both from the selection and combination schemes, the exponential smoothing models where the information criterion values could not be estimated or the lower or upper prediction interval of the furthest horizon was an outlying value, as determined by the interquartile range of the forecasts produced by the examined models.

Along with the results for the various selection and combination schemes, we offer the results for two simple benchmarks. The first is the aggregate selection of the best model by the evaluation criterion, referred as \textit{Aggregate-select}. The second is an equally weighted forecast combination, referred to as \textit{EQW-average}. Both of these benchmarks provide identical results for the two selection criteria considered.

\begin{table}[h]
	\centering
	\caption{The out-of-sample performance of the various selection and combination schemes.}\vspace{0.25cm}
	\begin{tabular}{cc|cccccc}
		Selection & \multirow{2}{1.4cm}{Scheme} & \multicolumn{2}{c}{Yearly} & \multicolumn{2}{c}{Quarterly} & \multicolumn{2}{c}{Monthly}\\
		Criterion && MASE & MSIS & MASE & MSIS & MASE & MSIS \\
		\hline
		BIC	& Aggregate-select & 3.512 & 45.524 & 1.192 &  9.953 & 0.988 & 8.893 \\
		& Criterion-select	& 3.412	& 33.175	& \textbf{1.166}	& \textbf{9.503}	& 0.949	& \textbf{8.175} \\
		& Precision-select	& 3.490	& 44.494	& 1.184	& 9.888	& 0.985	& 8.649 \\
		& Sensitivity-select	    & \textbf{3.309}	& \textbf{32.329}	& 1.174	& 10.368	& \textbf{0.948}	& 8.327 \\
		\cmidrule(lr){2-8}
		& EQW-average	    & 3.231	& 29.225	& 1.174	& 9.099	& 0.948	& 8.213 \\
		& Criterion-average	& 3.351	& 31.652	& 1.152	& 9.332	& 0.942	& 8.098 \\
		& Precision-average	& 3.247	& 29.935	& \textbf{1.147}	& \textbf{9.023}	& \textbf{0.916}	& \textbf{7.933} \\
		& Sensitivity-average	& \textbf{3.212}	& \textbf{29.180}	& 1.155	& 9.032	& 0.922	& 7.961 \\
		\hline
		Validation & Aggregate-select & 3.512 & 45.524 & 1.192  & \textbf{9.953} & 0.988 & 8.893\\	
		& Criterion-select	& \textbf{3.358}	& \textbf{38.937}	& 1.179	& 10.182	& \textbf{0.942}	& \textbf{8.640} \\
		& Precision-select	& 3.511	& 45.265	& 1.188	& 10.083	& 0.972	& 8.756 \\
		& Sensitivity-select	    & 3.374	& 39.119	& \textbf{1.178}	& 10.144	& 0.951	& 8.789 \\
		\cmidrule(lr){2-8}
		& EQW-average	    & 3.231	& 29.225	& 1.174	& 9.099	& 0.948	& 8.213 \\
		& Criterion-average	& 3.348	& 37.631	& 1.176	& 9.934	& 0.936	& 8.478 \\
		& Precision-average	& 3.251	& 30.004	& \textbf{1.159}	& \textbf{9.071}	& \textbf{0.925}	& \textbf{8.053} \\
		& Sensitivity-average	& \textbf{3.214}	& \textbf{29.149}	& 1.170	& 9.083	& 0.935	& 8.115 \\
		\hline
	\end{tabular}
	\label{tab:outofsample_performance}
\end{table}

We observe that, when a single model is selected, the Criterion-select scheme (regardless of the selection criterion) provides a reasonably good performance. Sensitivity-select is better than Criterion-select in the yearly frequency and when BIC is used. Also, Precision-select offers better performance with respect to prediction intervals for the quarterly data and the Validation selection criterion. However, it would be fair to say that there is not much to be gained from the contingency tables and the environmental information when a single model is selected. Overall, and as expected, Aggregate-select results in worse performance compared to other selection schemes, with the only exception being the MSIS for the quarterly frequency. 

The situation is different when models are combined. In this case, the Criterion-average scheme focuses on weights that have been estimated based on the information criteria values (for the BIC) or the validation performance of the models for a single time series, disregarding the general tendencies and performances of these models. On the other hand, Precision-select and Sensitivity-select take into account the performance of each model when applied to a large set of series. Sensitivity-average is the best approach for the yearly frequency, outperforming all other selection and combination approaches, including EQW-average. Similarly, Precision-average is the best approach for the seasonal (quarterly and monthly) frequencies. The gains from the application of Precision-average and Sensitivity-average are especially evident in the case of the performance of the prediction intervals. For example, the MSIS value for the Sensitivity-average at the yearly frequency is 7.8\% and 22.5\% lower than the respective values of Criterion-average for the BIC and Validation criteria, which, by turn, are lower than the respective Criterion-select values.

Comparing the results of Table \ref{tab:outofsample_performance} for the two selection criteria considered, BIC and Validation, we can generally notice small differences. However, BIC is overall better than Validation for the various selection and combination schemes. Regardless, the proposed Precision and Sensitivity selection/average schemes are by definition applicable for any selection criterion, as long as the respective contingency tables can be populated.

\begin{figure}[ht]
	\begin{center}
		\includegraphics[ width=6.4in ]{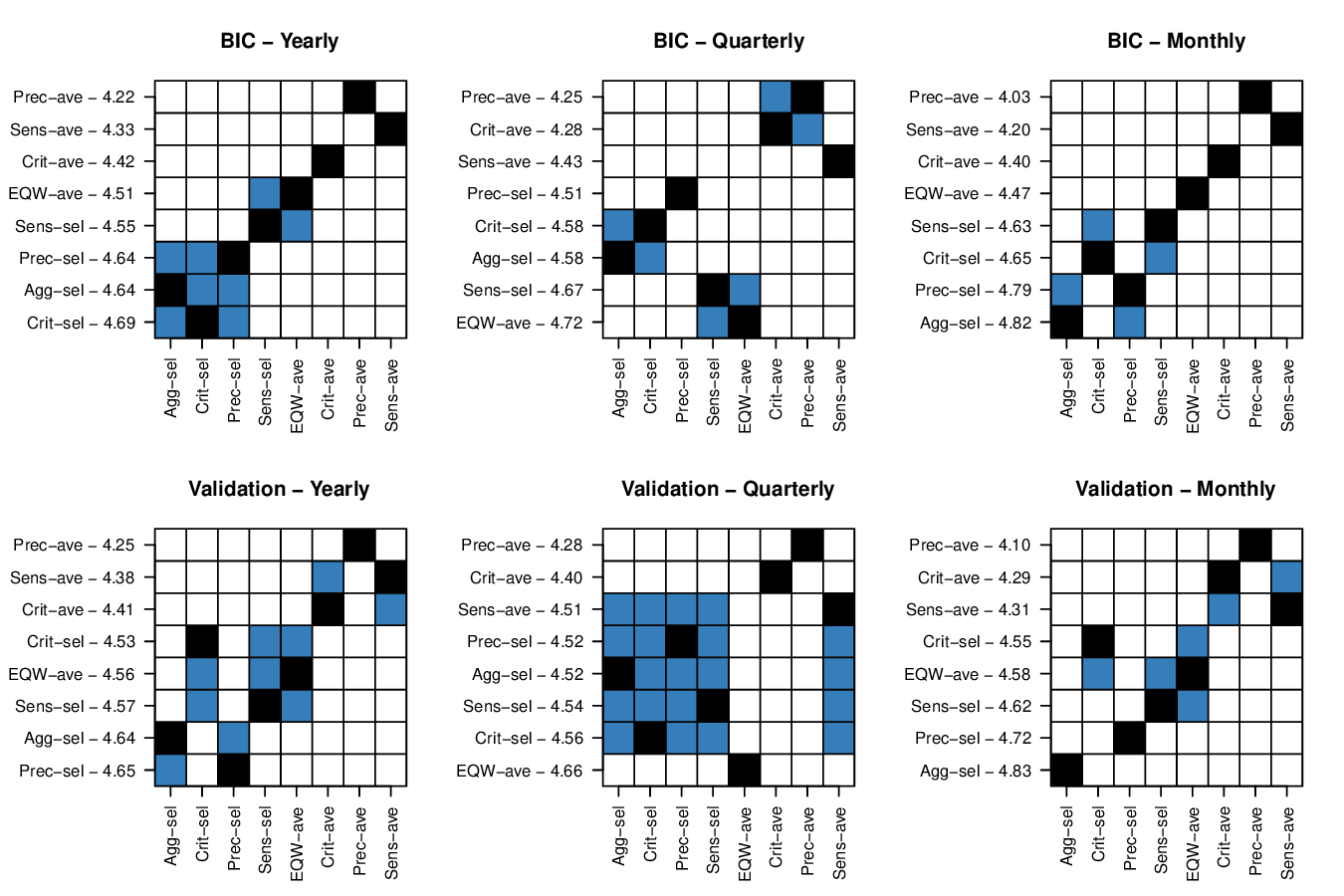}
		\caption{Nemenyi test results at a 5\% significance level for the MASE.}
		\label{fig:mcbmase}
	\end{center}
\end{figure}

\begin{figure}[ht]
	\begin{center}
		\includegraphics[ width=6.4in ]{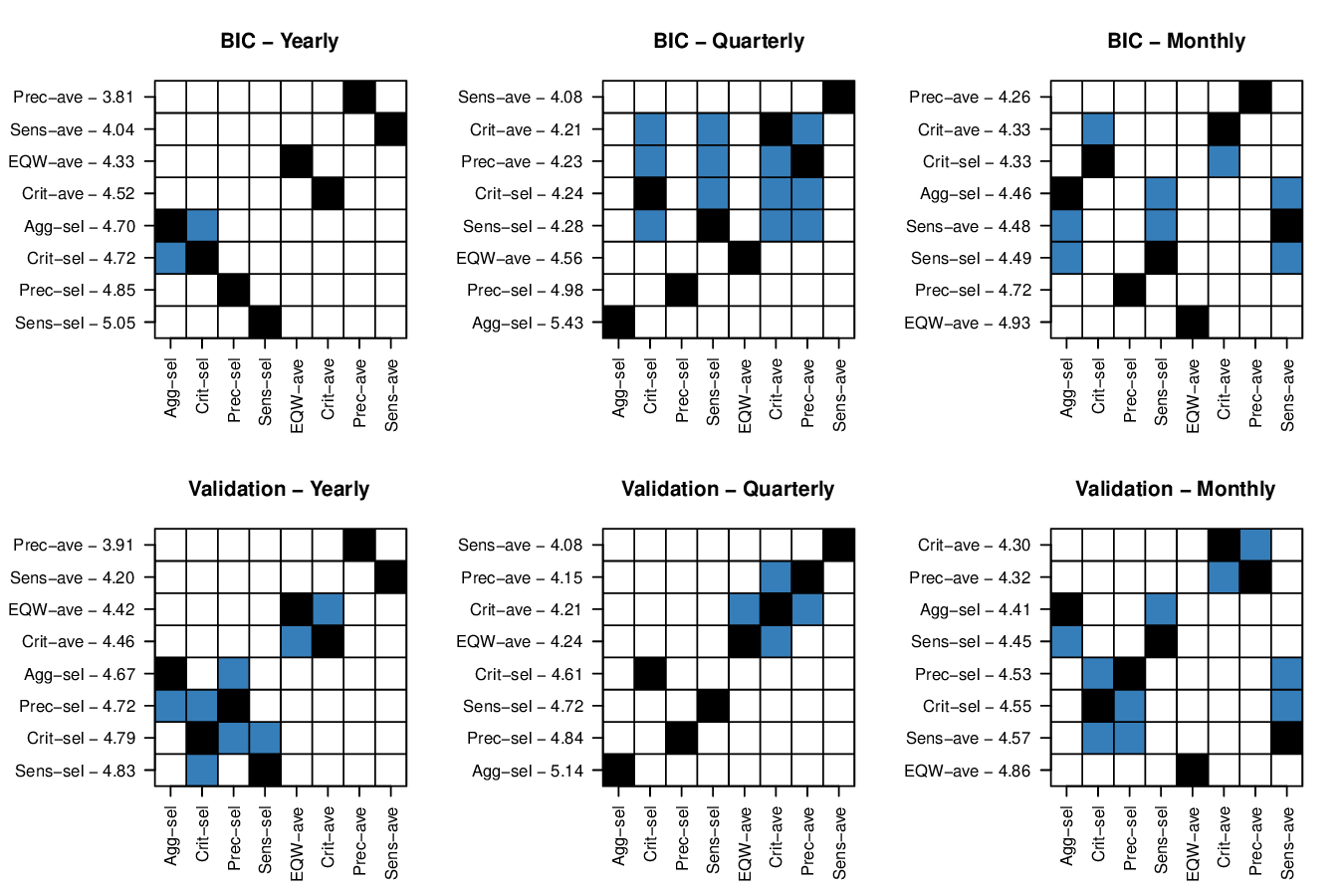}
		\caption{Nemenyi test results at a 5\% significance level for the MSIS.}
		\label{fig:mcbmsis}
	\end{center}
\end{figure}

Next, we perform nonparametric multiple comparisons using the Friedman and the post-hoc Nemenyi tests. The results from the application of these tests allow us to check whether or not the differences between the performance of the various selection and combination schemes are statistically significant. It is worth noting that these tests do not rely on distributional assumptions, while they focus on the ranked rather than the absolute performance of each scheme. We use the \texttt{nemenyi()} function of the \textit{tsutils} package for R. The significance results at a 5\% level are presented in figures \ref{fig:mcbmase} and \ref{fig:mcbmsis} for the MASE and the MSIS respectively. The considered schemes are presented from best (top row) to worst (bottom row) based on their average ranks. The columns' order follows the presentation of the schemes in table \ref{tab:outofsample_performance}. For each row, the black cell represents the scheme being tested; blue cells suggest that the scheme depicted in the row has an average rank that is not statistically different than the scheme in the respective column; and white cells suggest statistically significant differences. As an example, focusing on the first panel of Figure \ref{fig:mcbmase} (yearly data and BIC selection criterion), the equal-weighted average (``EQW-Ave''), which has an average rank of 4.51, is not statistically different, at a 5\% level, to Sensitivity-select (``Sens-Sel''), but it is statistically different to all other selection and combination schemes.

Three major observations arise from Figures \ref{fig:mcbmase} and \ref{fig:mcbmsis}. First, the Precision-average scheme is ranked always first in terms of MASE, regardless of the frequency of the data or the selection criterion. Moreover, it is statistically better than all other schemes, with the only exception being the quarterly data and the BIC criterion where there is no evidence of statistical different average ranks between Precision-average and Criterion-average. Second, the good performance of the Precision-average scheme is also evident in the yearly and monthly frequencies for the MSIS measure. However, the other combination scheme that utilises revised base-rate information, the Sensitivity-average, is significantly better than all others in the quarterly data. Third, there is no evidence that one of the selection schemes, Aggregate-select, Criterion-select, Precision-select, and Sensitivity-select, performs consistently better than the others.

\begin{figure}[ht]
	\begin{center}
		\includegraphics[ width=6in ]{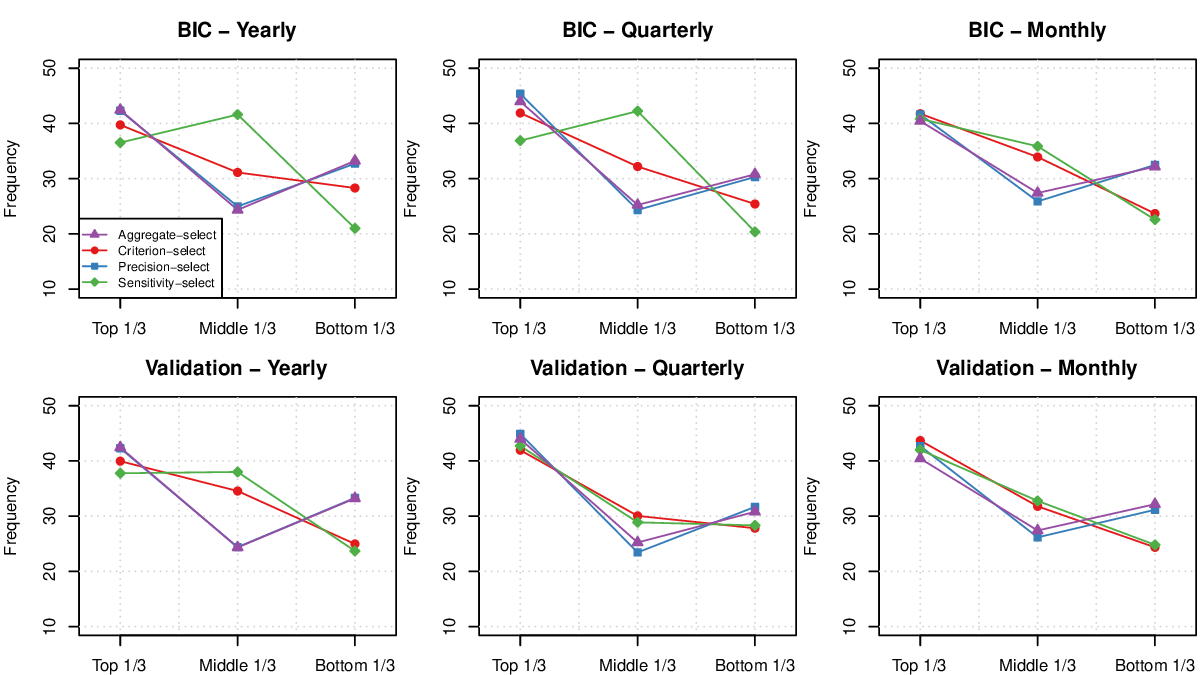}
		\caption{Selection frequencies of the top, middle, and bottom-ranked models for each selection scheme, analysed per data frequency and selection criterion.}
		\label{fig:ranks}
	\end{center}
\end{figure}

Next, we focus on the frequencies with which the four selection schemes opt for a model that is within the top, middle, or bottom third of the respective pool of available models. For example, given that the model pool consists of fifteen exponential smoothing models (six for the yearly data), then a scheme points to a model in the top 1/3 of the models when that model is ranked, based on its point forecast accuracy, in $[1, 5]$ (or $[1, 2]$ for the yearly data). Figure \ref{fig:ranks} presents the respective selection frequencies for each selection criterion and data frequency. We observe that Precision-select and Aggregate-select point to a model in the top-third more often than the other two selection schemes (Criterion-select and Sensitivity-select) for the yearly and quarterly data, irrespective of the selection criterion. However, Precision-select and Aggregate-select also opt more often than the other selection schemes a model that is ranked in the bottom-third of the models. Despite the similarities in the model ranks selected by Aggregate-select and Precision-select, the latter offers better overall performance, as observed in table \ref{tab:outofsample_performance}.

Another interesting observation arises from the ranked performance of the models selected by Sensitivity-select. In three of the six panels (BIC - Yearly, BIC - Quarterly, and Validation - Yearly), we see that Sensitivity-select opts significantly more frequently than the other two schemes for a model ranked in the middle-third and less frequently for models in either the top or bottom thirds. This suggests that Sensitivity-select selects less frequently the best models but also avoids more frequently the worst models. In that sense, Sensitivity-select works similarly to how humans select models \citep{Petropoulos2018-mt}. Sensitivity-select results in models with similar ranks to Criterion-select for the monthly data frequency but also for the quarterly data and the validation criterion.

\section{Discussion}\label{sec:discussion}

Our empirical results suggest that combining models using precision and sensitivity information significantly improves the performance of a system rather than focusing solely on the per-series information. Effectively, our results show that the base-rate information has a useful role to play in model selection and model combination. Suitably revising the base rates offers a forecasting performance that is superior to exclusively focusing on the case-specific information. Solely focusing on the base-rates, as showcased by the Aggregate-selection scheme, is not appropriate and a balance is needed between the environmental (aggregate) and individual (per series) information regarding the performance of a forecasting model.

Our approach is a very simple case of cross-learning, as the base-rate information is built by the application of the models within the pool on a large number of time series. Compared to other approaches that utilise cross-learning, our approach offers transparency while being intuitive. Normally, meta-learning and cross-learning approaches require extensive feature engineering \citep{Montero-Manso2020-et} and preprocessing or scaling of the data \citep{Kang2020-xq}. In contrast, our approach involves virtually zero ad-hoc modelling decisions or setting of hyperparameters, with the only exceptions being the choice of the pool of forecasting models and the selection of selection and evaluation criteria. It is widely automated and requires limited to none judgmental input from the modeller. As such, our approach on forecast combinations based on revised base-rates could be offered as part of automatic and batch forecasting solutions.

Similar to other cross-learning approaches, our approach consists of an offline and an online part. The offline part corresponds to the population of the contingency tables, while the online part corresponds to the use of these tables to estimate the revised base-rates. It is worth-mentioning that the additional calculations for the online part, once the values of the selection criteria have been estimated, are trivial and result in negligible additional computational cost. However, populating the contingency tables can be costly, even more for the validation selection criterion compared to information criteria such as the BIC. In any case, it would be reasonable to assume that, in a relatively constant environment, the offline part would not be updated in every review period (i.e., every time one needs to produce forecasts).

Usually, weighted-based combination approaches estimate a unique set of weights for each target series. An example is the Criterion-average approach, where the weights are estimated based on the values of the selection criterion for each model when applied to a particular (the target) series. However, this is not true for the Precision-average and Sensitivity-average approaches. We calculate only $K$ sets of weights for each of these approaches, with each set of weights being applied based on the model that is selected. In this sense, our combination weights are ``static'' given a reference set of series. It is not the first time that static combination weights are proposed in the literature. For instance, \cite{Collopy1992-my} proposed the use of static weights when combining between four models towards estimating levels and trends. However, contrary to them, our static weights are not arbitrarily selected but are directly linked with the environmental performance (base-rates) of the models. As such, the combination weights for Precision-average and Sensitivity-average will change if the set of reference series used for calculating the base-rate also changes.

One advantage of model combinations through revised base-rates is that they do not rely on specific selection or evaluation criteria, or even a standard pool of models. In this study, we focused on a single evaluation criterion (MAE) solely for purposes of brevity; the choice of MAE was made so that it links to the performance indicator used for measuring point-forecast accuracy (MASE). However, a different evaluation criterion would be more appropriate in other settings. Moreover, in our work we showed results for two selection criteria, BIC and Validation. The results are similar for other information criteria (such as AIC or AICc), while our approach could work with any other selection criterion, such as cross-validation. Finally, we limited our pool of models to exponential smoothing models. As long as the selection criteria values are comparable, then one could consider a pool that includes forecasting methods or models from several different families.

In our empirical design, we use suitable subseries of the target series to form the reference set of series and populate the contingency tables necessary for the Precision-based and Sensitivity-based schemes. We withheld an appropriate number of observations such that the evaluation criterion is calculated over a period that matches the required forecast horizon of the target series. The use of subseries of the target series inherently offers contingency tables that are representative to the target series. However, it would only work when the available series are many and long. Even in our case, we had to drop roughly 11\% of the series when populating the contingency tables for the validation criterion. In the case where the target series are short or the number of the target series is low, then the reference set should be a distinct set of series suitably selected such that it is representative to the target set. Approaches for measuring the representativeness and diversity of sets of series are offered by \cite{Kang2017-ha} and \cite{Spiliotis2020-gl}.

In this study, we used a large set of real data pooled from three major forecasting competitions, M, M3 and M4. We should highlight that our results are based on relatively low frequency data (monthly to yearly) but we have no reason to believe that the insights gained cannot be generalised to other, higher frequency data. Also, we would like to mention that we do not intend to directly compare the achieved performances presented in this paper with any of the original submissions in the aforementioned forecasting competitions. While we did not use the test data explicitly, having access to the hold-out data renders any comparison with the competitions' participants unfair.

\section{Concluding remarks}\label{sec:conclusions}

In this study, we argued that the selection of models for time series forecasting should not exclusively focus on the values of selection criteria applied on each series individually, but the base-rate information should also be taken into account, i.e., how often a particular model performs best on the out-of-sample. We argued that such ``environmental'' information of the performance of the various models should be revised with the case-specific information towards obtaining probabilities that each of the candidate models is indeed the correct one. Such probabilities can then be used for model (forecast) selection or forecast averaging. Our approach is a very simple case of cross-learning, while also being in-line with the agenda of Bayesian inference through loss functions.

Our empirical analysis was based on the point-forecast accuracy and performance of the prediction intervals using a large set of real-life time series. Our results showed that combination approaches based on Precision and Sensitivity information can outperform both individual and aggregate selection or combination, while conceptually being in-between the two. In some cases, the differences in performance were statistically significant. The insights gained were similar for the two selection criteria (BIC and validation), the various sampling frequencies, and the two measures (MASE and MSIS) considered.

Future research could focus on context-specific data and how contingency tables can be populated to better suit the needs of organisations with a small and relatively uniform sets of data. Moreover, in this paper we limited our attention to exponential smoothing models. As our approach is not structurally limited to these models, it would be interesting to see how it performs when selecting and combining over a more diverse pool of models. A final promising avenue of future research will be to see whether the algorithms proposed can be extended beyond forecasting, to model combination for estimating common parameters across models, in the spirit of \citep{LavRoc2016}.

\bibliographystyle{elsarticle-harv}

\bibliography{refaggr}
\end{document}